\documentclass[12pt]{article}

\usepackage{amsmath}				
\usepackage{amsfonts} 	
\usepackage{setspace}				
\usepackage{graphicx}				
\usepackage{wrapfig}				
\usepackage{times}					

\usepackage[top=1in, bottom=1in, left=1in, right=1in]{geometry} 	
\usepackage[compact]{titlesec}								

\usepackage[font=footnotesize,labelfont=bf,labelsep=period,width=0.75\textwidth]{caption} 	
\usepackage[font=footnotesize,labelfont=bf,labelsep=period]{subcaption} 				
\DeclareCaptionSubType*[arabic]{figure} 										
\DeclareCaptionLabelFormat{subfiglabel}{Figure #2} 								
\usepackage{ifthen}
\usepackage[dvipsnames]{xcolor}
\usepackage[normalem]{ulem}

\newboolean{showcomments}
\newboolean{showedits}
\newboolean{showtodo}

\begin{document}

\begin{center}

{\Large Berkeley-Leibniz Relativity and Symplectic Flag Manifolds}\par\vskip12pt

{\normalsize B. E. Eichinger}\par
    
{\small\itshape Department of Chemistry, University of Washington, Seattle, Washington 98195-1700}\par


\end{center}

\begin{abstract}

The relativity of space and motion arguments of Berkeley, Leibniz, and others, provide an alternative to the absolute space of Newton, and are a basis for describing the physics of many-particle systems.  The general philosophical ideas are made concrete by showing that points in a configuration space of generalized coordinates can be moved by the action of a group.  By interpreting the way in which a group acts on two or more systems of points that are at first isolated and then brought into contact, it is shown that interactions can be represented by the action of cosets of the group.  Application of this principle to $n$ fermions leads to the proposal that the symplectic group, $Sp(n)$, is the largest group of isometries of the configuration space of physical particles, and that interactions between particles, carried by bosons, are represented by cosets of the group.  The manifold of interactions between $n$ fermions is represented by the complete quaternionic flag variety $Sp(n)/ \bigotimes _1^n Sp(1)$.  Some aspects of the geometry of these spaces are developed and several contacts with physics are explored, primarily by means of the Lie algebra of $Sp(n)$. 

\end{abstract}

\section*{Introduction}
Newton formulated his three laws of mechanics in the context of an \emph{absolute} three dimensional Euclidean space.  In addition to the three coordinates that locate objects in this flat space, the first law states that momentum space is also Euclidean.  And Newtonian time is strictly linear.  This seven-dimensional ideal, flat, Newtonian geometry has been the foundation of the physical sciences.  For example, Maxwell's equations were formulated in the Newtonian era, and are explicitly dependent on absolute space and time coordinates.  It was not until the Special Theory of Relativity was devised that absolute space and time was superseded.  However, because the Special Theory is formulated to be consistent with Maxwell's equations, it necessarily retains a vestige of absolute space and time -- the inertial frame can be understood to be the device that welds the space-time frame of Special Relativity onto the absolute Newtonian frame of Maxwell's equations.   

Not long after Newton's laws were formulated Berkeley and Leibniz objected to absolute space on philosophical grounds.  Berkeley asserted, for example, that an isolated body cannot be moved, and thus one cannot define its velocity.  This is not the place to delve into the philosophies of Berkeley\cite{Berkeley} and Leibniz,\cite{Alexander} and subsequently Kant, Mach and others\cite{Alexander}. Let it suffice to say that recognition of the relativity of space and motion has a long history.  It was not until Einstein included time in relativistic considerations that the Berkeley-Leibniz philosophy became a reality for physics.  However, as noted above, in building Special Relativity atop Maxwell's equations, the introduction of an inertial frame is required and this is not wholly consistent with the Berkeley-Leibniz (BL) philosophy, which would rather shun an inertial frame.\footnote{It is interesting to interpret Newton's laws in terms of modern manifold theory and Berkeley-Leibniz relativity to show that the three laws  are not restricted to flat absolute space. The first law states that there is a tangent plane at a material point of a manifold. The second law identifies curvature at a material point with force.  The first two laws are very general statements that hold for any reasonable (smooth) manifold. The third law equates the curvature at $A$ due to $B$ to have the same magnitude as the curvature at $B$ due to $A$, and this balance of forces is perfectly consistent with Berkeley's ideas.  The forces acting within an isolated system sum to zero, and since a truly isolated system has never experienced an outside force, it never accelerated and therefore is not moving, just as Berkeley opined.}

The BL argument brings to mind the elementary geometrical notion that $N$ points in general position define an $N-1$ dimensional space.  Berkeley expressed this by noting that ``. . . in every motion it be necessary to conceive \emph{more bodies than one} (my emphasis) . . .''\cite{Berkeley}  The apparently limitless choice of spatial dimension that accompanies the elementary linear independence of $N$ points raises the first of several issues that must be addressed in attempting to produce a practical theory while following the BL philosophy.  What limits space to three dimensions?  The Special Theory of Relativity requires inclusion of the time dimension as well.  Does this require many time coordinates?  The answer to the latter question appears to be yes: The Bethe-Salpeter equation is a two-body relativistic quantum theory using two sets of space-time variables, \emph{i.e.}, \underline{two time variables}, to describe the motions of two interacting fast particles moving relative to an inertial frame.\cite{BetheSal}  

The Special Theory dispensed with absolute space and time, but extension to many bodies appears to require a multiple time formulation that is not easily interpreted.  This difficulty has to be addressed in a theory describing the relative motions of many fast objects while respecting the Special Theory in its original intent -- it applies to the relative rectilinear motion of two bodies, one of which defines an inertial frame.  By confronting the many-body problem from the BL perspective, one might hope to shed light on the meaning of many time dimensions, as exemplified by the Bethe-Salpeter theory. 

While the General Theory of Relativity relates gravitation to the curvature of a non-Euclidean geometry, the formulation relies on a flat space at infinity.  This is built into the theory as a boundary condition -- sufficiently far from a gravitating body the Special Theory should hold.  The assumption that space is asymptotically flat motivates use of the Ricci tensor rather than the Riemann tensor to characterize the curvature.  This flat boundary condition at infinity, which has been inherited from Newton, permeates virtually all of science, despite the obvious fact that it is an ideal that cannot be experimentally verified.  

Clearly, space itself is not observable.  The only observable is the relation between objects.  We infer the properties of space and time from these relations.  To construct a theory without invoking a preconceived geometry it will be necessary to take a non-traditional route to formulate the relations between objects.  It is not possible to discuss relations between objects in the context of physical laws that are formulated in terms of traditional coordinates.  A different approach is required.

While considering how a many-body description of the relative positions of objects might be expressed without specifying an intrinsic space, a group action comes to mind.  For example, a permutation group can act on a set of objects without reference to any additional external structure.  But geometry is crucial for physical theory and objects must be allowed to move continuously relative to one another, implying that the group action has to be continuous.  Here is a mathematical setting that might be used to begin consideration of relative motions in a general context, without explicit dependence on a particular underlying fixed manifold structure.  Now the questions are: Can a purely mathematical description be made to conform to known physical facts?  How much experimental information is required to fix upon a physically rational group?  And given this structure, what can be derived from it?  These questions will be addressed in the following.

\section*{Constructing the Group Action}
Consider a set $a$ of points with relative coordinates $x_a$, over the real $\mathbb{R}$, complex $\mathbb{C}$, or quaternion $\mathbb{H}$ rings, the coordinates being unspecified except to say that they are sufficient to describe the geometrical relations between the points.  Associated with this set of points will be a function, $\psi_a(x_a)$, which might be scalar, vector or tensor valued, and which  will enable us to describe \emph{physical properties} of the set of points.  Define a group $G(a)$ that acts transitively on the coordinates such that $g(a)\in G(a)$ is represented by the continuous right action $g(a):\psi_a(x_a) \to R_{g(a)}\psi_a(x_a)=\psi_a(x_{a}g(a))$.\cite{Folland}  Transitive action in this context is understood to encompass all physically acceptable configurations of the set of points, which implies that transitivity will be defined by an acceptable manifold structure.  

Now consider a second set $b$ of points, utterly independent of the set $a$, and subject to the similar action $R_{g(b)}\psi_b(x_b)=\psi_b(x_{b}g(b))$.  The physical content of asserting that the two sets of points are independent is that they do not interact with one another.  The groups $G(a)$ and $G(b)$ will have different dimensions if the sets $a$ and $b$ do not have the same number of points.  However, there is no intrinsic difference between $a$ and $b$, so that $G(a)$ and $G(b)$ belong to the same class of groups.  

Now allow the two sets to be merged into one, $ab:=a\cup b$, such that $R_{g(ab)}\psi_{ab}(x_{ab})=\psi_{ab}(x_{ab}g(ab))$, while conserving the number of points.  The combined group, $G(ab)$, moves all the points in the merged set, while $G(a)$ moves the $a$ set independently of $b$ and \emph{vice versa}.  The elements of the coset $G(ab)/G(a)\times G(b)$ move $a$ because of the presence of $b$ while simultaneously moving $b$ because of the presence of $a$.  A change in a system that is bought about by the presence of another system signifies that there is a physical interaction between them.  The coset represents this interaction!  
  
This argument can be applied recursively to any number of systems, $(a:z):=a\cup b\cup c\cup \cdots \cup z$, with the group $G(a:z)$ moving everything, while the coset $G(a:z)/G(a)\times G(b)\times\cdots\times G(z)$ is understood to be the part of $G(a:z)$ that moves what are now subsystems relative to one another owing to their mutual interactions, while $G(a)\times G(b)\times\cdots\times G(z)$ moves the parts of each subsystem independent of the other subsystems.  

The argument also extends downward.  Consider a single point -- by itself it cannot move, but it can move relative to its neighbors in a set of points.  Changing notation to count points, the interaction between a single point and $n$ others will be represented by $G(n+1)/G(1)\times G(n)$. This is a principal bundle,\cite{K&N} with a $G(1)$ fiber sitting on each point.  Elaborating on this notation pursuant to the discussion in the previous paragraph, and remembering that the ring $\mathbb F$ might be $\mathbb R,\mathbb C$, or $\mathbb H$, the general coset (bundle) structure of interest for a system consisting of $m$ subsystems becomes $G(n,\mathbb F)/\bigotimes_{j=1}^m G(n_j,\mathbb F)$, $\sum_{j=1}^m n_j=n$.  This structure is our candidate to represent the interactions between any number of subsystems.  A singled-out point has a symmetry group $G(1,\mathbb F)$.  If we were dealing with simple mathematical points, one might be inclined to set $G(1,\mathbb F)=1$, since a point by itself cannot move.  However, we are trying to describe real objects in this BL program, so this is not a good choice.  To fix the ring we need to attach physical properties to the mathematical points, and that brings the functions $\psi_a(x_a)$ into play.   

\subsection*{Specifying the Field}   
The continuous transformation group, $G(n,\mathbb F)$, that is under construction is a Lie group.  The enormous descriptive advantage of a Lie group is that it has an algebra of infinitesimal generators acting on vector-valued functions.  This motivates looking for a symmetry group acting on functions, particularly those for single particles, that is compatible with the natural action of our group acting on coordinates.  What properties must functions have to be acceptable to describe fermions?  Many fermions have mass and charge, but all of them have spin.  And spin is the only property that can be defined with nothing other than a group.  Particle theory recommends that a representation of $G(1,\mathbb F)$ act on a two-component spinor, $\psi_1(x)$. The additional reason for choosing a spinor is that the class of functions 
\begin{equation}\label{induce}
\psi_1(x\xi)=\sigma(\xi^{-1})\psi_1(x),
\end{equation}
with $\sigma$ a unitary representation of $\xi\in G(1,\mathbb F)$ is a natural construction in induced representation theory.\cite{Folland}  Functions of this class will enable us to construct linear combinations of states, as will be seen as the subject develops.  

Returning to general considerations, two systems that are not interacting with one another will be subject to the left action (or a right action, in makes no difference to the argument)
\begin{equation}\label{redrep}
\left[{\begin{array}{*{20}c}
 L_{g(a)} & 0  \\
 0 &L_{g(b)} \\
\end{array}}\right]
\left[{\begin{array}{*{20}c}
 \psi_{a}(x_a) \\
 \psi_{b}(x_b) \\
\end{array}}\right]=\left[{\begin{array}{*{20}c}
 \psi_{a}(g(a)^{-1}x_a) \\
 \psi_{b}(g(b)^{-1}x_b) \\
\end{array}}\right].
\end{equation}
Within the context of a larger group, diag$(g(a),g(b)) \in G(ab)$ is a reduced representation, which corresponds to non-interacting subsystems.  

Two or more systems that might be considered as independent of one another, to whatever desired accuracy, will be represented by a block diagonal structure as exemplified by eq. (\ref {redrep}).  As the systems are brought into contact, the dimension of the vector space $[\psi_{a},\psi_{b}]$ should not be allowed to change.  (Particle creation and annihilation will be considered later.)  It will be convenient to track the vector dimension, even though the particles comprising the systems might exchange, such that 
\[
L_{g(ab)}\left[{\begin{array}{*{20}c}
 \psi_{a}(x_{ab}) \\
 \psi_{b}(x_{ab}) \\
\end{array}}\right]=\left[{\begin{array}{*{20}c}
 \psi_{a}(g(ab)^{-1}x_{ab}) \\
 \psi_{b}(g(ab)^{-1}x_{ab}) \\
\end{array}}\right].
\]
With this background we can now get to the specifics of the group structure.

The candidates for our spin group in eq. (\ref{induce}), $\sigma(\xi^{-1})\sim G(1,\mathbb F)$, are $O(3), U(2)$, and $Sp(1)$, as these have isomorphic algebras $\mathfrak {so}(3)\sim \mathfrak {su}(2)\sim\mathfrak{sp}(1)$ (temporarily ignoring the abelian component in the unitary case). Because we want a family of groups, applicable to any $n$, it is natural to look to the four classical families: $A_n, B_n, C_n$, and $D_n$.\cite{Helgason}  Which family -- orthogonal, unitary, or symplectic -- with corresponding rings $\mathbb{R}$, $\mathbb{C}$, or $\mathbb{H}$, should we choose?  

The simplest answer that applies to systems of any $n$ and is compatible with our coset structure requires that we select from $O(3n) \sim Spin(3n)$, $U(2n)$, and $Sp(n)$.  The real dimensions of the groups are\cite{Helgason}
\begin{align*}
\textrm{dim}[O(3n)] = {} &3n(3n-1)/2 \\
\textrm{dim}[U(2n)] = {} &4n^2 \\
\textrm{dim}[Sp(n)] = {} &n(2n+1).
\end{align*}
One would first like to understand a single particle, with candidate cosets $G(n+1,\mathbb F)/G(1,\mathbb F)\times G(n,\mathbb F)$ describing the interactions between a particle and its surroundings.  The real dimensions of the coset spaces appropriate to this structure are\cite{Helgason}
\begin{align*}
\dim[O(3n+3) /O(3)\times O(3n)] = {} &9n, \\
\dim[ U(2n+2)/U(2)\times U(2n)] = {} &8n, \\
\dim [Sp(n+1) /Sp(1)\times Sp(n)] = {} &4n.
\end{align*}
That is, the interaction between the subject particle and each particle in the surroundings has real dimension 9, 8, or 4.  Pairwise interactions with eight or nine degrees of freedom pose a significant interpretation problem, but those with four might be interpreted as natural space-time coordinates.  The only realistic choice is the symplectic group. By the recursion argument above, single particles are the ultimate subsystems, so that the coset space $Sp(n)/\bigotimes_1^n Sp(1) \sim Sp(n)/Sp(1)^n$ describes interactions between particles.  This is a complete quaternionic flag manifold\cite{Mare} -- it has a rich mathematical structure.    The physical significance of the geometry of these manifolds will emerge as the subject develops.

Structures similar to this have been encountered many times.  For example, Goldstone bosons have a $G/H$ interpretation.\cite{Weinberg}.   A structure based on $G(1,\mathbb C)=U(1)$ and $U(n)/U(1)^n$ was discussed by Atiyah\cite{Atiyah,A&B} in relation to a paper by Berry and Robbins.\cite{B&R}  Atiyah's unitary version provides an avenue to become acquainted with our subject matter by use of a commutative algebra.  However, the unitary model takes the single particle states to be simple complex functions, and spin has to be attached separately.  The $Sp(1)^n$ fiber structure is more comprehensive.

There is a geometrical bonus in the quaternion division algebra -- it provides algebraic rigidity to space-time, which answers one of the questions posed in the Introduction.  To illustrate difficulties with the real case, one might ask:  Where are the $O(2n)/O(2) \times O(2n-2)$ or $O(4n)/O(4) \times O(4n-4)$ particles?  And in the group product, how does one interpret the fact that coordinates get mixed up in physically unattractive ways?  This difficulty applies equally to non-compact coset spaces such as $SO(n,3n)/[SO(1,3)]^n$ that are motivated by Special Relativity; these are also excluded from consideration.  With quaternions such questions do not arise -- quaternions provide a natural, self-contained, space-time structure that is maintained through all products, and these contain just the usual scalar and vector products of $\mathbb R^3$, as will be seen later.  Briefly, a quaternion $q$ is defined by four real numbers, $x_0,x_1,x_2,x_3$, and a basis, ${\bf e,i,j,k}$, the basis having one commuting element, ${\bf e}$, and three anti-commuting elements, ${\bf i,j,k}$, with $q=x_0{\bf e}+x_1{\bf i}+x_2{\bf j}+x_3{\bf k}$; the conjugate quaternion is $\bar{q}= x_0{\bf e}-x_1{\bf i}-x_2{\bf j}-x_3{\bf k}$.

The symplectic group is also denoted by $U\left({n,\mathbb{H}}\right)$, \emph{i.e.}, it is a unitary group over the quaternion ring $\mathbb{H}$.  The group is both complex symplectic and unitary: $Sp(n)\sim Sp(2n,\mathbb{C})\cap U(2n,\mathbb{C})$.\cite{Simon}  The representation in the $Sp(2n,\mathbb{C})$ form will be addressed later.  The group $Sp(n)$ is compact and is not manifestly covariant in the usual sense.  However, the coset spaces are projective spaces, and they have the \emph{appearance} of being non-compact, as will be seen when we develop their metrics.  The relation to the Lorentz group and deSitter and Anti-deSitter (AdS) spaces will be established as our subject develops.

\section* {Composite Systems}

Individual subsystems are distinguishable when they are isolated from one another.  An isolated subsystem consisting of $k$ fermions is represented by a vector space $\psi_k$ that transforms under the left or right action of $Sp(k)$ as described above.  In the context of the group $Sp(n), n> k$, representing a large number of subsystems, the interactions between the subsystems can be described by the partial flag variety $Sp(n)/Sp(k_1)\times Sp(k_2)\times \cdots\times Sp(k_m)$, with $\sum_{j=1}^{m}k_j=n$.  This flexibility of group representations provides a graceful transition from a subsystem of indistinguishable particles, where invariance under a permutation group is implicit in $Sp(k_{i})$, to preserving the individuality, $i\ne j$, of subsystems.  As the theory develops, the representations of $Sp(n)$ that are induced by representations of the "maximal factor group" $\bigotimes_{j=1}^{m}Sp(k_j)\sim H_{\mathbb K}$,  ($\mathbb K$ is a partition of $n$) will surely be of interest.  

In the mathematical literature, flag manifolds are most often constructed from $GL(n,\mathbb {F})$ by moding out a maximal parabolic subgroup, $G/\mathcal{P}$.\cite{Fulton} The choice of $U(n,\mathbb H)$ with maximal factor group used here provides a more cogent physical interpretation than this alternative representation, which is not to say that $GL(n,\mathbb H)/\mathcal{P}$ might not be well suited to understand some aspects of the theory.
  
\section* {The Action of $SL\left(2,\mathbb{C}\right)/\{\pm\mathbf {1}\}\sim SO\left(3,1\right) $}

The isomorphism of groups in this section title has been exploited in physics, but usually as a sidelight.  Now the isomorphism becomes important to an understanding the relation between BL relativity and the Special Theory.  Our focus in developing BL relativity is to understand the action of groups: A group is the \emph{a priori} structure, not $\mathbb{R}^3$ or the light cone.  The arguments presented thus far identify a quaternion as the natural representation of the spatial relation, and interaction, between two particles.  Knowing just this, and ignoring the many-body setting that has been developed, we might ask for the largest group that preserves the norm (magnitude or strength) of a single quaternion, $q$.  For this purpose use is made of the well-known representation of quaternions by Pauli matrices modulo $\sqrt{-1}$.  A quaternion is isomorphic to the matrix, $m(\mathbb{C}^2):=\mathbb{R}^{+}\times SU(2)= \mathbb{R}^{+}\times Sp(2,\mathbb{C}) $, representation.  The largest group that preserves the determinant $\sim$ norm in this representation, is $u\in SL\left(2,\mathbb{C}\right)/\{\pm\mathbf {1}\}$ acting by a similarity transformation $u: m \to u^\ast mu$ ($u^\ast$ is the transposed conjugate of $u$).  The well-known group isomorphism $SL(2,\mathbb{C})/\{\pm\mathbf {1}\}\sim SO(3,1) $ is all that is needed to map between the quaternion representation and space-time with a Lorentz signature.  To state the matter in a somewhat different way: Suppose we are given the $SL(2,\mathbb C)$ action on a quaternion, but we \emph{insist} on a \underline{real} interpretation of the coordinates.  The psychological imperative of real coordinates forces one to accept the $SO(3,1)$ interpretation.  This is one way to make contact with the Special Theory, but there are others as will be seen.

\section* {Two Aspects of Time}
All practical measures of time are based on cyclic motion.  Whether time is measured with an atomic clock or by the number of earthly trips around the sun, cyclic motions set the scale.  Linear time, as expressed by Galilean, Newtonian, or cosmological time, has the character of a topological winding number.  Notwithstanding macroscopic classical motions with a beginning and an end, the physically important temporal functions of electrodynamics and quantum theory are cyclic functions of the form $\exp(i\omega t)$, where $t$ is Galilean time.  Natural motion is cyclic, which implies that every such motion constitutes a cyclic clock.  Cyclic motion is pervasive, occurring on length scales from atoms to planets to galaxies.  
 
One of the four real coordinates that comprise a quaternion has to be identified as a cyclic temporal variable.  Every interaction has its own cyclic time coordinate,\footnote{This provides a rationale for the two times that appear in the Bethe-Salpeter equation,\cite{BetheSal} where each electron moves relative to the inertial frame and each requires a temporal variable.  It is also noted that BL relativity eliminates the twin paradox, as neither Bob nor Ray, separated in an otherwise empty universe but for their individual life support systems, is privileged to an inertial frame.} and this coordinate is completely interchangeable with the spatial variables by means of elementary rotations of the quaternions (or the action of $SL(2,\mathbb C)$ in the restricted sense above). Cosmological (winding number) time is not subject to the same transformation.  We will see later that there is a natural and unique place to insert linear time into the theory, and all cyclic spacetime coordinates will be parametrically determined by the linear (winding number) time. We rely on Galilean time to time-order events, on all length scales, and this will be built into the theory at the appropriate place.

\section* {Bottom up Construction and Group Action}

The complete quaternionic flag variety $Sp(n)/Sp(1)^n$  is difficult to handle directly, so a (local) parameterization via the factorization
\begin{equation}\label{product}
Sp_n/\bigotimes_1^n Sp_1 \sim [Sp_n/Sp_1\times Sp_{n-1}]\times[Sp_{n-1}/Sp_1\times Sp_{n-2}]\times\cdots\times[Sp_2/Sp_1\times Sp_1]
\end{equation}
enables one to build up solutions by solving the smallest problems first.  It is convenient for this section, and many others to follow, to use the shortened notation $Sp_k:=Sp(k)$.   This parameterization is a product of rank one hyper-K\"ahler manifolds,\cite{Swann,HKLR} which are also spheres: $S^{4k}=Sp_{k+1}/Sp_{1}\times Sp_{k}$.  At the bottom of this nest of coset spaces is $Sp\left(2\right)/Sp\left(1\right)\times Sp\left(1\right)$, which is the space of solutions of the Yang-Mills functional\cite{At}.  Using standard group isomorphisms this space is alternatively represented by\cite{Lawson} 
\[
Sp(2)/Sp(1)\times Sp(1)\sim SO(5)/SO(3)\times SO(3) \sim SO(5)/SO(4) \sim S^4,
\]
\emph{i.e.}, the four-sphere\cite{At,Lawson}.  This  space of interactions between two particles is identified as an instanton.  More succinctly stated, instantons represent bosons.  The four-sphere will be discussed at greater length later.  

The action of the respective groups on their coset spaces is given by linear fractional transformations. To show this, embed the coset space $Sp_{k+1}/Sp_1\times Sp_k$ in $Sp_n$, $n>{k+1}$, and let it be parameterized by the elements 
\[
\exp \left[{\begin{array}{*{20}c}
 0 & 0 & 0 \\
 0 & 0 & {\xi} \\
 0 & {-\xi^{\ast}} & 0 \\
\end{array}} \right]
= \left [{\begin{array}{*{20}c}
 1 & 0 & 0 \\
 0 & (1-ZZ^{\ast})^{1/2} & Z \\
 0 & {-Z^{\ast}} & ( 1-Z^{\ast}Z)^{1/2} 
\end{array}} \right]
\]
derived from the Lie algebra\cite{Gilmore}.  Here $\xi$ is a $k$-dimensional vector over the quaternions, $\xi^\ast$ is the conjugate transpose, $\xi^\ast=\bar{\xi}'$, and 
\begin{align*}
Z = (\xi\xi^{\ast})^{-1/2}[\sin(\xi\xi^{\ast})^{1/2}]\xi = {}& \xi (\xi^{\ast}\xi)^{-1/2}\sin(\xi^{\ast}\xi)^{1/2},\\
(1-ZZ^{\ast})^{1/2}={}&\cos(\xi\xi^{\ast})^{1/2},\\
(1-Z^{\ast}Z)^{1/2}={}&\cos(\xi^{\ast}\xi)^{1/2},
\end{align*}
with the trigonometric functions being defined by their formal power series.   For ${k+1}<n$ the above representation is embedded in the $n\times n$ larger matrix as shown; in the sequel this embedding will be understood and the padding will be omitted.  Alternatively, for the purposes of this section, one might think about the properties of a system of $k+1$ fermions in isolation, ignoring any potential interactions with the surroundings.

An element $g\in Sp_{k+1}$, expressed in a conforming partitioning, is  
\[
g=\left[{\begin{array}{*{20}c}
 A & B \\
 C & D
 \end{array}}\right] ;  \quad g^{-1} = g^* = \left[{\begin{array}{*{20}c}
  A^* & C^* \\
  B^* & D^*
  \end{array}}\right],
 \]
and this acts by
\[
gxH=yH
\]
\[
gxH=\left[{\begin{array}{*{20}c}
 A & B \\
 C & D
 \end{array}}\right]
 \left [{\begin{array}{*{20}c}
 \left( {1-ZZ^*}\right)^{1/2} & Z \\
 {-Z^*} & \left( {1-Z^*Z}\right)^{1/2} 
\end{array}} \right]H
\]
\[
yH=\left [{\begin{array}{*{20}c}
 {A\left( {1-ZZ^*}\right)^{1/2}-BZ^*} & {AZ+B\left(1-Z^*Z\right)^{1/2}} \\
 {C\left( {1-ZZ^*}\right)^{1/2}-DZ^*} & {CZ+D\left( {1-Z^*Z}\right)^{1/2}} 
\end{array}} \right]H
\]
where $H:= Sp_1\times Sp_k$.  This construction works equally well for a larger class of coset spaces than is implied here; $\xi$ might be a $j\times k$ matrix of quaternions.  Then $Z$ belongs to $Sp(j+k)/Sp( j)\times Sp(k)$, which provides compatible partitioning for the action of $Sp(j+k)$.  This larger coset space will be used in the development.  

Physically what is happening is that the system, the $Sp(j)\in H$ part, and the surroundings, the $Sp(k)\in H$ part, may experience arbitrary changes in their spin/internal states as a result of the action of $H$, independent of any interaction between the system and surroundings.  The system and surroundings each consist of parts which act upon one another -- those within the system act on one another independent of the surroundings and \emph{vice versa}.  For an isolated particle this internal motion is along the $Sp(1)$ fiber.  All of the internal degrees of freedom are encompassed by $H$.  We want to know how the coset space behaves under the general action of $g$, and this requires that the action of $H$ be eliminated by taking ratios.  The action of $G\sim Sp(j+k) $ on the Grassmannian coordinates of $G/H\sim Sp(j+k) /Sp( j)\times Sp(k)$, $(g: X\to Y)$,  is\cite{Hua}
\begin{equation} \label{lft} 
Y =  (AX+B)(CX+D)^{-1} 
 = (-XB^*+A^*)^{-1}(XD^*-C^*)
 \end{equation}
with $X=Z(1-Z^*Z)^{-1/2}=(1-ZZ^*)^{-1/2}Z$.  The space $X$ is the Grassmannian.  It has the appearance of being non-compact, as a point $X$ where $A^*-XB^*$ is singular is mapped to infinity by the group action.  An alternative route to formulating the action of the group on the Grassmannian goes through the Stiefel manifold, $Sp(j+k)/Sp(k)$, consisting of $j\times (j+k)$ dimensional matrices $U$ of quaternions such that $UU^\ast = 1$.\cite{K&N}  

\section* {Induced Representations}

It is probable that induced representation theory will be important in constructing representations.  The structure of induced representations is a perfect expression of the invariance of the inner product of wave functions under the gauge group.  The construction begins by `averaging' over the fiber with a fixed cross-section (coset space).  Let $H$ be a maximal factor group (structure group),\cite{Sternberg} \emph {e.g.}, $Sp(k)\times Sp(n-k)$, of $Sp(n)$, with $\xi\in H$ and $\sigma(\xi)$ a unitary representation of $H$.  A function defined by 
\begin{equation*}
f_{\alpha}(x)=\int_{H}\sigma(\eta)\alpha(x\eta)d\eta
\end{equation*}
has been averaged over all internal motions of subsystems, with a fixed interaction between subsystems.  Here $\alpha(x)$ is a map from $Sp(n)$ to a Hilbert space with an inner product.  It is easy to show that $f_{\alpha}(x\xi)=\sigma(\xi^{-1})f(x)$, as in eq. (\ref{induce}).\cite{Folland,Simon}  Furthermore, since $\sigma$ is a unitary gauge group the inner product $<f,f>$ is independent of $H$.  This construction enables one to construct linear combinations of states, and in particular, to construct linear combinations of eigenvectors of the action of $G$.   

\section*{Metric and Curvature}

This section is standard material and only the results will be presented\cite{Hua}.  The invariant line element on $Sp( j+k) /Sp( j)\times Sp(k)$ is given by 
\begin{align}
ds^2 = {} & \textrm{tr}\left[\left({1+XX^*}\right)^{-1}dX\left({1+X^*X}\right)^{-1}dX^*\right] \label {metrica}\\
  = {} & \textrm{tr}\left[\left({1+XX^*}\right)^{-1}dXdX^*-\left({1+XX^*}\right)^{-1}dXX^*\left({1+XX^*}\right)^{-1}XdX^*\right] \label{metricb}
\end{align}
where the second version follows from $\left({1+X^*X}\right)^{-1}=1-X^*\left({1+XX^*}\right)^{-1}X$; $X$ is a $j\times k$ matrix of quaternions.  One can show that the metric is invariant to an appropriately defined inversion.  In the simplest case, $Sp_{2}/Sp^{2}_{1}$, the metric is invariant to $X\to X^{-1}$.

	A number of projective invariants might also be constructed.  Let $X_a$, resp. $Y_a$, denote particular points in the Grassmannian manifold, such that $g:X_a \to Y_a$ for $g\in G$.  Using eq. (\ref {lft}) it is easy to show that 
\begin{align*}
1+Y_{a}Y^{\ast}_b=&{}(-X_{a}B^\ast + A^\ast)^{-1}(1+X_{a}X^{\ast}_b)(-BX^{\ast}_b + A)^{-1},\\
1+Y^{\ast}_{a}Y_{b}=&{}(X^{\ast}_{a}C^\ast + D^\ast)^{-1}(1+X^{\ast}_{a}X_b)(CX_b + D)^{-1},\\
Y_a-Y_b=&{}(-X_{a}B^\ast + A^\ast)^{-1}(X_a-X_b)(CX_b + D)^{-1},\\
=&{}(-X_{b}B^\ast + A^\ast)^{-1}(X_a-X_b)(CX_a + D)^{-1}.
\end{align*}
The surprising last equality can be verified by direct calculation from eq. (\ref{lft}) using the left and right members of that equation for $Y_a$ and $Y_b$ in two different ways.  From these a variety of invariants and cross-ratios can be constructed.  For example, it is easy to show that the cross ration of four points, 
$\textrm{tr}[(Y_a-Y_b)(Y_c-Y_b)^{-1}(Y_c-Y_d)(Y_a-Y_d)^{-1}]$, is invariant under the action of the group.  The permutations $Y_c-Y_b\to Y_b-Y_c$ simultaneously with $Y_a-Y_d \to Y_d-Y_a$ is a reflection of the surprising equality.  This set of equations can be used to derive the metric in eq. (\ref{metrica}).
  
A different way of looking at the linear fractional transformation reveals the relation between the group and the vector space on which it operates.   The origin, $X=0$ in eq. (\ref{lft}), is mapped to $Y=BD^{-1}=-(A^{*})^{-1}C^{*}$ by the group, which gives the coordinates of the Grassmannian $Y$ in terms of the coordinates in the group manifold.  ($A$ and $D$ are continuously connected to the identity, so their inverses exist in a neighborhood of the identity.) Since $gg^{*}=g^{*}g=1$ it follows that 
\begin{align}\label{A}
(1+YY^{*})={}&(AA^{*})^{-1},\\
(1+Y^{*}Y)={}&(DD^{*})^{-1}.
\end{align}
The exterior derivative of $Y=B/D$ is
\begin{equation}\label{dY}
dY=dBD^{-1}-BD^{-1}dDD^{-1}=dBD^{-1}+(A^{*})^{-1}C^{*}dDD^{-1}=(A^{*})^{-1}\omega_{12}D^{-1}
\end{equation}
where 
\begin{equation}\label{omdef}
\omega=g^{*}dg=\left[\begin{matrix}
A^{*}&C^{*}\\
B^{*}&D^{*}
\end{matrix}\right]\left[\begin{matrix}
dA & dB\\
dC & dD
\end{matrix}\right]=\left[\begin{matrix}
\omega_{11}&\omega_{12}\\
\omega_{21}&\omega_{22}
\end{matrix}\right]
\end{equation}
Using eqs. (\ref{A}-\ref{omdef}) it is easy to show that eq. (\ref{metrica}) is alternatively written as 
\begin{equation}\label{metric2}
ds^2=\textrm {Tr}(\omega_{12}\omega^{*}_{12}).
\end{equation}

Define the (quaternion-valued) basis on which $G= Sp(j+k)$ acts on the right to be ${\bf e}=({\bf e}_{j}, {\bf e}_{k})$.  Let ${\bf e}^0$ be the basis at the identity; then ${\bf e}={\bf e}^{0}g$ and $d{\bf e}={\bf e}^{0}dg = {\bf e}g^{-1}dg= {\bf e}g^{*}dg={\bf e}\omega$.  This assumes that $G$ acts transitively on the basis.  The Maurer-Cartan form,\cite{Chern,Chern1946} $d\omega+\omega\wedge\omega$, with $\omega$ defined in eq. (\ref{omdef}), is
\begin{equation}\label{MC}
d\omega+\omega\wedge\omega=\left[\begin{matrix}
d\omega_{11}+\omega_{11}\wedge\omega_{11}+\omega_{12}\wedge\omega_{21}&d\omega_{12}+\omega_{11}\wedge\omega_{12}+\omega_{12}\wedge\omega_{22}\\
d\omega_{21}+\omega_{21}\wedge\omega_{11}+\omega_{22}\wedge\omega_{21}&d\omega_{22}+\omega_{21}\wedge\omega_{12}+\omega_{22}\wedge\omega_{22}
\end{matrix}\right],
\end{equation}
but this is identically zero because $d\omega = d(g^{-1}dg)=-g^{-1}dgg^{-1}\wedge dg=-\omega\wedge\omega$.  Each block in eq. (\ref{MC}) must vanish.  In particular, the diagonal blocks are $d\omega_{11}+\omega_{11}\wedge\omega_{11}+\omega_{12}\wedge\omega_{21}=0$ and $d\omega_{22}+\omega_{22}\wedge\omega_{22}+\omega_{21}\wedge\omega_{12}=0$.  But by the definition of the curvature tensor, $d\omega+\omega\wedge\omega-\Omega=0$, this gives two parts for the curvature tensor $\Omega$: $\Omega_{11}=-\omega_{12}\wedge\omega_{21}$ and $\Omega_{22}=-\omega_{21}\wedge\omega_{12}$, consistent with the treatment of Chern.\cite{Chern1946} The off-diagonal blocks in  eq. (\ref{MC}) give $d\Omega_{aa}=[\Omega_{aa},\omega_{aa}]$, which is the second Bianchi identity.  Calculations similar to this can be found in Chpt. 8 of the Ref. \cite{Sharpe} where the Maurer-Cartan equations for projective spaces $\mathbb{P}^n(\mathbb {R})$ are considered in a form similar to this block matrix method.  An alternative derivation, beginning with the equation for geodesics, is given by Wong.{\cite{Wong1}  The version presented here has the advantage of demonstrating that the tensor has two pieces, one piece being the curvature at system 1 due to the presence of system 2 and \emph{vice versa} for the second piece, which echoes Newton's third law.  From the Maurer-Cartan form it is clear is that if $d\omega_{aa}+\omega_{aa}\wedge\omega_{aa}=0$, which is what the diagonal elements become for $G=Sp(j)\times Sp(k)$, \emph{i.e.}, if $G$ is restricted to the isotropy subgroup $H$, the curvature tensor vanishes.  Viewed in another way, a reducible representation signifies that subsystems are isolated from one another, the curvature tensor vanishes, and no forces act between the subsystems.  This is an important demonstration of self-consistency.

From here it is straightforward to simplify the curvature tensor.  Since $\omega$ is skew-symmetric, $\omega_{21}=-\omega^{*}_{12}$, giving $\Omega_{11}=\omega_{12}\wedge\omega^{*}_{12}$ and $\Omega_{22}=\omega_{21}\wedge\omega^{*}_{21}=\omega^{*}_{12}\wedge \omega_{12}$.  Now look back to eq. (\ref{metric2}) to see how the curvature is related to the metric.  To render this in traditional form one may use eq. (\ref{dY}) to yield  
\[
\Omega_{11}=\omega_{12}\wedge\omega^{*}_{12}=A^{*}dYD\wedge D^{*}dY^{*}A=A^{*}dY(1+Y^{*}Y)^{-1}\wedge dY^{*}A
\]
and 
\[
\Omega_{22}=\omega_{21}\wedge\omega^{*}_{21}=D^{*}dY^{*}A\wedge A^{*}dYD =D^{*}dY^{*}(1+YY^{*})^{-1}\wedge dYD
\]
The traces of the two pieces of $\Omega$, i.e., $R_{aa}=\textrm{Tr}(\Omega_{aa})$, give 
\begin{align*}\label{Ricci}
R_{11}={}&\textrm{tr}[dY(1+Y^{*}Y)^{-1}\wedge dY^{*}(1+YY^{*})^{-1}]\\
R_{22}={}&\textrm{tr}[dY^{*}(1+YY^{*})^{-1}\wedge dY(1+Y^{*}Y)^{-1}]
\end{align*}
which is sufficient to show that the Ricci tensor is equal to the metric tensor.  The tensor has two pieces that have the same magnitude, which succinctly expresses Newton's third law.  The matrix dimensions of the two pieces are compatible with the dimensions of the two subsystems.  This identifies the $Sp_{j+k}/Sp_j\times Sp_k$ coset space as an Einstein space.\cite{Besse}  

The exterior products of the two one-forms that are required in the two parts of the curvature tensor are computed with the rules that the product of basis elements are standard, but the product of the differential coefficients is skew-symmetric.  The product of individual terms is $dx{\bf {e}_r}\wedge dy{\bf{e}_s}=(dx\wedge dy){\bf e}_r{\bf e}_s=-(dy\wedge dx){\bf e}_r{\bf e}_s=(dy\wedge dx){\bf e}_s{\bf e}_r, \lbrace {\bf e}_r,{\bf e}_s\rbrace=\lbrace {\bf i,j,k}\rbrace=\lbrace{\bf e}_1,{\bf e}_2,{\bf e}_3\rbrace$.  For the $Sp(2)/Sp(1)^2$ case, set $dY = dy_0{\bf e}_0+dy_1{\bf e}_1+dy_2{\bf e}_2+dy_3{\bf e}_3=dy_0{\bf e}_0+d{\bf y}$, using vector notation.  It follows that 
\begin{align*}
dY\wedge dY^\ast = {}&(dy_0{\bf e}_0+d{\bf y})\wedge(dy_0{\bf e}_0-d{\bf y})\\
={}&-2(dy_0\wedge d{\bf y}+d{\bf y}\wedge d{\bf y})\\
dY^\ast \wedge dY={}&(dy_0{\bf e}_0-d{\bf y})\wedge (dy_0{\bf e}_0+d{\bf y})\\
={}&2(dy_0\wedge d{\bf y}-d{\bf y}\wedge d{\bf y}).
\end{align*}
with
\[
dy_0\wedge d{\bf y}\pm d{\bf y}\wedge d{\bf y}=(dx_0\wedge dx_1\pm dx_2\wedge dx_3) {\bf e}_1+(dx_0\wedge dx_2\pm dx_3\wedge dx_1) {\bf e}_2+(dx_0\wedge dx_3\pm dx_1\wedge dx_2) {\bf e}_3.
\]
These are self-dual ($dY\wedge dY^\ast$) and anti-self-dual ($dY^\ast\wedge dY$) sectors.\cite{At}  The two components of the curvature tensor are alternatively interpreted as time-reversal symmetric, which symmetry is also apparent in the Lie algebra, $\mathfrak g = -\mathfrak g^\ast$.  Elements $g_{ij}\in \mathfrak g$ and $g_{ji}$ differ only in the signs of the scalar components.  The import of this symmetry will become apparent in subsequent sections.

\section*{Equation of Motion and Lie Algebra}

On returning to the wave function description, let  
\begin{equation}\label{eqmot}
\Psi\left(xH^{-1}\right)=\left[{\begin{array}{cc}
\psi_{\alpha}(xH^{-1}) \\
\psi_{b}(xH^{-1})
\end{array}}\right]=\sigma(H)\left[{\begin{array}{cc}
\psi_{\alpha}(x) \\
\psi_{b}(x)
\end{array}}\right]=\left[{\begin{array}{cc}
\sigma(h_{\alpha})\psi_{\alpha}(x) \\
\sigma(h_{b})\psi_{b}(x)
\end{array}}\right]
\end{equation}
be a square-integrable vector-valued function ($\mathbb H$-module) that is compatible with the matrix representation of the group that we have been working with.  Notation has been changed with the isotropy subgroups being replaced: $\textrm{diag}(g_1,g_2)\to H$, so as to be compatible with standard mathematical usage.  Here $x$ are coordinates of the coset, which was identified above as a Grassmannian.  The second version on the right is the natural generalization of eq. (\ref{induce}), and the last version is appropriate owing to the block diagonal structure of $H$.  (Our wave function is a vector bundle associated to the principle bundle.\cite{K&N,Darling})  It is useful to think of $\psi_{\alpha}(x)$ as the wave function of the system, and $\psi_b(x)$ as that of the surroundings.  The reason for the peculiar indexing, mixing Greek and Latin letters, will be apparent shortly. 

The most natural equation of motion that is consistent with quantum theory and our group is provided by a one-parameter local group action with Lie algebra $\mathfrak{g}$, so that
\begin{equation} \label {motion}
\partial \Psi/\partial t = \mathfrak{g}\Psi
\end{equation}
where $\mathfrak{g}$ consists of the infinitesmal generators of the Lie algebra, \emph{i.e.}, the Lie derivative.  The parameter $t$ is identified with Galilean (winding number) time because we want to use this equation to time-order events.  (We are using natural units with $\hbar=c=1$.  Furthermore, skew-symmetry of the Lie algebra, with concomitant suppression of $\sqrt{-1}$, is preferred over the Hermitian option so as to avoid clutter.)  This equation might be derived from a conservation principle; the total time derivative of a \textit{p}-form over a manifold is given by an equation of this type.\cite{Frankel}  Since $\mathfrak{g}^\ast = -\mathfrak{g}$, it also follows that $\partial (\Psi^\ast \Psi)/\partial t=0$, signifying that $<\Psi,\Psi>$ for an isolated system is conserved for all time.

We now need to fix a representation for the quaternions so as to present the Lie algebra.  The standard basis for the quaternions consists of $\{\mathbf{e,i,j,k}\}$, with 
\[
\mathbf{e}={\bf1}; \quad \mathbf{i}^2=\mathbf{j}^2=\mathbf{k}^2=-\mathbf{e}=-{\bf1}; \quad \mathbf{ij}=\mathbf{k}, \quad \mathbf{jk}=\mathbf{i},\quad \mathbf{ki}=\mathbf{j}
\]
A quaternion $q$ will be written $q:=w\mathbf{e}+x\mathbf{i}+y\mathbf{j}+z\mathbf{k}$, with conjugate $\bar{q}:=w\mathbf{e}-x\mathbf{i}-y\mathbf{j}-z\mathbf{k}$.  (Note that conjugation is the parity operator, and $-\bar{q}$ is cyclic time reversal.)  The norm $|q|$  of $q$ is defined by $|q|^2\mathbf{e}=q\bar{q}= \bar{q}q=(w^2+x^2+y^2+z^2)\mathbf{e}$.  The derivative is most naturally defined such that $dq/dq = 1$, which implies that 
\[
d/dq=\frac{1}{4}\left(\mathbf{e}\partial/\partial{w}- \mathbf{i}\partial/\partial{x}-\mathbf{j}\partial/\partial{y}-\mathbf{k}\partial/\partial{z}\right).
\]
Choose a matrix representation of a quaternion in a basis of Pauli matrices (modulo $\sqrt{-1}$ ) as 
\[
q = \left[{\begin{array}{cc}
{w+iz} & {x + iy} \\
{-\left(x-iy\right)} & {w-iz}
\end{array}}\right]=\left[{\begin{array}{cc}
{\rho_1} & {\rho_2} \\
{-\bar\rho_2} & {\bar\rho_1}
\end{array}}\right]
\]
and  
\begin{equation}\label{quat}
q^*=\bar{q}^{\prime}=\left[{\begin{array}{cc}
{\bar\rho_1} & {\bar\rho_2} \\
{-\rho_2} & {\rho_1}
\end{array}}\right]^{\prime}=\left[{\begin{array}{cc}
{\bar\rho_1} & {-\rho_2} \\
{\bar\rho_2} & {\rho_1}
\end{array}}\right]
\end{equation}
where $\bar{r}$ is the complex conjugate and $r^{\prime}$ is the transpose of the matrix $r$ . This is the $m(\mathbb {C}^2)$ representation mentioned above.  The differential operator in the matrix representation is 
\begin{equation}\label{deriv}
d/dq=\frac{1}{2}\left[{\begin{array}{cc}
{\partial/\partial{\rho_1}} & {-\partial/\partial{\bar\rho_2}} \\
{\partial/\partial{\rho_2}} & {\partial/\partial{\bar\rho_1}}
\end{array}}\right]
\end{equation}
This choice for $d/dq$ gives $dq/dq = 1$ and $d\bar{q}/dq =0$, but $dq^\ast/dq = -1/2$.  Note that this derivative is defined to act in the sense of quaternion multiplication -- there is an implicit sum over row and column indices.  It is sometimes easier to work with the algebraic representation, $\mathbb{H}$, than with the matrix representation in explicit calculations. However, quaternionic differentiations are intrinsically non-commutative and can be hazardous  -- for calculation of commutation relations, for example, it is best to use another convention for differentiations.  

Having shown the structure of the quaternionic derivative, it will be temporarily discarded in favor of one that is more naturally suited to component-wise differentiation in the matrix representation.  (The author finds that this gives better control over the summation convention.) Define $Q = (\zeta_{\alpha a}); 1\le \alpha \le 2k, 1\le a \le 2(n-k)$ to be a matrix of quaternions in the $m(\mathbb {C}^2)$ representation, with 
\[
\left[\begin{matrix} 
    \zeta_{2\mu - 1,2t-1} & \zeta_{2\mu - 1,2t} \\
     \zeta_{2\mu ,2t-1} & \zeta_{2\mu,2t}\\
   \end{matrix}\right]
=\left[\begin{matrix} 
      z^{(1)} & z^{(2)} \\
      -\bar z^{(2)} & \bar z^{(1)}\\
   \end{matrix}\right]_{\mu t}=q_{\mu t}; 1\le\mu\le k, 1\le t \le (n-k).
\]
The differential operator $\partial/\partial \zeta_{\alpha a}$ is now defined such that $\partial \zeta_{\alpha a}/\partial \zeta_{\beta b}=\partial_{\beta b}\zeta_{\alpha a} =\delta_{\alpha\beta}\delta_{ab}$.  As an $m(\mathbb {C}^2)$ operator, this is
\[
\partial_{\mu t}=\left[\begin{matrix}
\partial/\partial z^{(1)} & \partial/\partial z^{(2)}\\
-\partial/\partial \bar{z}^{(2)} & \partial/\partial \bar{z}^{(1)}\\
\end{matrix}\right]_{\mu t},
\]
which is just the transpose of the operator in eq. (\ref{deriv}).  More simply stated, for the purpose of developing the infinitesimal generators of the Lie algebra, $Q$ is being regarded as a matrix of complex variables and their conjugates.  But because the $m(\mathbb {C}^2)$ representation contains both $\zeta$ and $\bar\zeta$, there is an additional `almost complex' structure.  Define the matrix 
\[
j=\left[\begin{matrix} 
   0 & 1 \\
   -1 & 0\\
   \end{matrix}\right], 
\]
such that the transformation 
\[
j'q j=\left[\begin{matrix} 
   0 & -1 \\
   1 & 0\\
   \end{matrix}\right]\left[\begin{matrix}
      z_1 & z_2 \\
      -\bar z_2 & \bar z_1\\
   \end{matrix}\right]\left[\begin{matrix} 
   0 & 1 \\
   -1 & 0\\
   \end{matrix}\right]=\left[\begin{matrix} 
      \bar z_1 & \bar z_2 \\
      - z_2 & z_1\\
   \end{matrix}\right]=\bar {q}.
\]
converts an $m(\mathbb {C}^2)$ representation to its complex conjugate.  (Note that $j$ acts in the $m(\mathbb {C}^2)$ representation in the same way that the quaternion basis element $\bf{i}$ acts on a quaternion.  The nomenclature for the `almost complex' action is traditional in the mathematical literature.)  The transformation facilitates differentiation of conjugate quaternions (the summation convention will now be used): 
\[
\partial \bar \zeta_{\alpha a}/\partial \zeta_{\beta b}=\partial_{\beta b}\bar\zeta_{\alpha a}=\partial_{\beta b}(J'_{\alpha \gamma}\zeta_{\gamma c}J_{ca})=J'_{\alpha\beta}J'_{ab}=J_{\beta\alpha}J_{ba}.  
\]
In the following the short-hand notation $J'QJ\to \bar Q$ will be used: it is understood that the pre- and post-$J$ factors are of the form ${\bf 1}\otimes j$ with appropriate dimension of the unit matrix, ${\bf 1}$, to be compatible with the $2k\times2(n-k)$-dimensional matrix $Q$.

The infinitesimal generators of the Lie algebra are:
\begin{align}
h_{\alpha \beta}&=  \zeta_{\alpha b}\partial_{\beta b}-\bar \zeta_{\beta b}\bar\partial_{\alpha b}\label{h_op}\\ 
H_{ab} & = \zeta_{\mu a}\partial_{\mu b}-\bar \zeta_{\mu b}\bar\partial_{\mu a}\label{Hop}\\
p_{\alpha a}&=   \bar\partial_{\alpha a}+ \zeta_{\alpha b}\zeta_{\mu a}\partial_{\mu b}\\
&=(\delta_{\alpha\beta}+\zeta_{\alpha b}\bar \zeta_{\beta b})\bar \partial_{\beta a}+\zeta_{\alpha b}H_{ab}\label{p_op2}\\
&=(\delta_{ab}+\zeta_{\mu a}\bar \zeta_{\mu b})\bar \partial_{\alpha b}+\zeta_{\mu a}h_{\alpha\mu}\label{p_op1}.
\end{align}
It is easy to see that $h^{*}=-h$ and $H^{*}=-H$.  Furthermore, $\bar p$ can be understood to differ from $p$ by quaternion conjugation (action of the parity operator).  The infinitesimal generators are written more succinctly as 
\begin{align*}
h&=Q\partial'-(Q\partial')^\ast\\
H&=Q'\partial-(Q'\partial)^\ast\\
p&=(1+QQ^\ast)\bar\partial+QH';
\end{align*}
other condensed versions of $p$ can also be written down.

All of the generators of the Lie algebra of $Sp(n)$ are parameterized by the Grassmannian coordinates of $Sp(n)/Sp(k)\times Sp(n-k)$.  Expressions for the individual operators are dependent on the given partitioning, and will be different for a different values of $k$ with fixed $n$.  

The generators satisfy the following commutation relations:
\begin{align}
[h_{\alpha \beta},h_{\mu \nu}]={}&\delta_{\beta \mu}h_{\alpha \nu}-\delta_{\alpha \nu}h_{\mu\beta}-J_{\beta \nu}(hJ)_{\alpha \mu}+J_{\mu \alpha}(Jh)_{\beta \nu}\\
[H_{ab},H_{cd}]={}&\delta_{bc}H_{ad}-\delta_{ad}H_{cb}-J_{bd}(HJ)_{ac}+J_{ca}(JH)_{db}\\
[h_{\alpha \beta},H_{ab}]={}&0\\
[p_{\alpha a},h_{\mu \nu}]={}&-\delta_{\alpha \nu}p_{\mu a}-J_{\alpha \mu }(Jp)_{\nu a}\\
[p_{\alpha a},H_{bc}]={}&-\delta_{ac}p_{\alpha b}+J_{ab}(pJ)_{\alpha c}\\
[p_{\alpha a},p_{\beta b}]={}&-J_{ab}(hJ)_{\alpha \beta}-J_{\alpha \beta }(HJ)_{ab}\\
[\bar p_{\alpha a},p_{\beta b}]={}&\delta_{\alpha \beta}H_{ba}+\delta_{ab}h_{\beta \alpha}
\end{align}
There are many different ways of writing these equations.  For example, $(Jh)_ {\alpha \beta }= (JhJ'J)_ {\alpha \beta }=(\bar h J)_ {\alpha \beta }=-(h'J)_ {\alpha \beta }=-(J'h)_ {\beta\alpha}=(Jh)_ {\beta\alpha}$.   To work out these somewhat tedious commutation relations, it is helpful to construct intermediate operators, such as $T_{ab} = \zeta_{\mu a}\partial_{\mu b}$, and use the rule $[xy,z]=x[y,z]+[x,z]y$ repeatedly.  Using $T_{ab}$, it follows that $H_{ab}=T_{ab}-\bar T_{ba}=T_{ab}-J'_{bc}T_{cd}J_{db}$;  using matrix notation, this is just $H=T-(J'TJ)'=T-J'T'J$.  From this form of the generator it is easy to see that  $JH=JT-T'J$, so that $(JH)'=T'J'-J'T=-T'J+JT=JH$.  From the commutators above others can be constructed by taking conjugates and making use of the symmetries of the generators. 
	 
It is convenient to simplify notation for the following discussion to explore one aspect of the theory.  Define a vector $(v,V)^{\prime}$ with components in $\mathbb{H}$, where $\dim(v)=k$ and $\dim(V)=n-k$.  Previously, our state vector, $\psi_1(x_1)$, for a single particle in the fundamental  representation of $m(\mathbb C^2)$ might have been regarded as $\psi_1(x_1)=(\phi^{(1)},\phi^{(2)})' $, with $\phi^{(\nu)}\in \mathbb C$.  On constructing this state vector, there exists a "conjugate" state $(-\bar{\phi}_{2},\bar{\phi}_{1})$ that is immediately constructible by virtue of the quaternionic structure of $Sp(n)$.  In either the $m(\mathbb C^2)$ or $Sp(1)$ picture, consider a higher dimensional irreducible representation. (This might be constructed from tensor products of the fundamental representation, for example.)  The Lie algebra acts on this representation vector space by 
\begin{equation}\label{transitions}
\left[{\begin{array}{cc}
{\mathfrak h_v} &{- \mathfrak p} \\
{\mathfrak p^\ast} & {\mathfrak h_V}
\end{array}}\right]\left[{\begin{array}{cc}
{v}\\
{V}
\end{array}}\right]=\left[{\begin{array}{cc}
{\mathfrak h_{v}v-\mathfrak {p}V}\\
{\mathfrak p^*v+\mathfrak h_{V}V}
\end{array}}\right],
\end{equation}
where $\mathfrak h_v{\in}Sp_k$ and $\mathfrak h_V{\in}Sp_{n-k}$.  The $\mathfrak h_v$ and $\mathfrak h_V$ operators "rotate" the system ($Sp_k$ part) and the surroundings ($Sp_{n-k}$ part), respectively, while the $\mathfrak p$ and $\mathfrak p^*$ operators induce transitions between them.  

Transitions induced by $\mathfrak p$ and $\mathfrak p^\ast$are easily seen using standard operations.  Suspend the summation convention, and let $V_a$ be an eigenvector with eigenvalue $n_a$ of $h_a$ in the Cartan subalgegra of $Sp_{n-k}$.  The action of $p_{{\alpha}{ a}}$ on this eigenvector is easily deduced from  
\begin{equation}
h_{a}p_{{\alpha}{ a}}{V_a}= p_{{\alpha}{ a}}h_{a}{V_a}+[h_{a},p_{{\alpha}{ a}}]V_a=n_{a}p_{{\alpha}{ a}}{V_a}+p_{{\alpha}{ a}}{V_a}=(n_{a}+1)p_{{\alpha}{ a}}{V_a}
\end{equation}
with use of the commutation relations.  Simultaneously, the conjugate $\bar p_{{\alpha}{ a}}$ acts on $v_\alpha$, an eigenvector of $h_\alpha \in Sp_k$ with eigenvalue $n_\alpha$.  This is a lowering operator, as is seen from 
\begin{equation}
h_{\alpha}\bar p_{{\alpha}{ a}}v_\alpha=\bar p_{{\alpha}{ a}}h_{\alpha}v_\alpha+[h_{\alpha},\bar p_{{\alpha}{ a}}]v_\alpha=n_{\alpha}\bar p_{{\alpha}{ a}}v_\alpha-\bar p_{{\alpha}{ a}}v_\alpha=(n_{\alpha}-1)\bar p_{{\alpha}{ a}}v_\alpha.
\end{equation}
The location of raising and lowering operators in this construction is arbitrary as are the signs; interchanging $p$ and $\bar p$ reverses the direction of the transition, raising $v_\alpha$ and lowering $V_a$.  This demonstrates that the algebra encompasses exchange of excitations between system and surroundings in either direction.  

Energy exchange will likely be more interesting to study in the context of the group, where topological considerations will be important. The theory has the right ingredients for application of the "bubbling off" theorem\cite{Lawson}, and a study of excitation transfer in that context might provide insight into the relation between localization of curvature and conical intersections\cite{BerWilk} of cross sections.  

The Laplace-Beltrami (LB) operator, $ \Delta$, associated to these operators is the trace of the square of the matrix of generators (to within a sign).  Thus
\begin{equation}\label{lap}
\Delta = \textrm{tr}\left(\mathfrak{h}_1\mathfrak{h}_1^*+\mathfrak{p}\mathfrak{p}^*\right)+\textrm{tr}\left(\mathfrak{h}_k\mathfrak{h}_k^*+\mathfrak{p}^*\mathfrak{p}\right).
\end{equation} 
This is only one among many of the composite operators that can be formed from the generators of the Lie algebra.  

\subsection* {Geodesics}
The formal solution of  eq. (\ref{motion}) is 
\begin{equation}\label{soln}
\Psi(t,xH) = \exp(t\mathfrak g)\Psi(0,xH).
\end{equation}
It is important to note that all geodesics on the group are of the form $\exp(t\mathfrak g)$.\cite{Simon,Price}  Another interesting aspect of eq. (\ref{soln}) is that winding-number time reversal and spatial inversion (quaternion conjugation) is the identity operation, since $t\mathfrak g \to -t\mathfrak g^\ast = t\mathfrak g$ because $\mathfrak g^\ast = -\mathfrak g$.  Define $g(t)=\exp(t\mathfrak g)$ to be the current value of $g\in G$.  It follows that $g(t)=g(t)[g(t_0)]^{-1}g(t_0)=g(t)g(-t_0)g(t_0)=g(t-t_0)g(t_0)$, so that eq. (\ref{soln}) can also be written $\Psi(t,xH) = g(t-t_0)\Psi(t_0,xH)$.  One may start the clock from any given state.  The practical implication of this is that proofs that rely on $G$ being near the identity can be used throughout, provided $G(t-t_0)$ acts on the current state.  For a more global perspective, the reader may want to consider eq. (\ref{soln} when interpreted with $t=$ cosmological time.

The eigenvalues of a Lie group lie in the maximal torus, and for a symplectic Lie algebra are pure imaginary  and occur in conjugate pairs.\cite{Simon,Fulton}  It is possible to write eq. (\ref{soln}) in terms of the eigenvalues of $\mathfrak g$, which is likely to be useful in looking at stationary states for small systems.  But there is another feature of eq. (\ref{motion}) that needs elaboration, and this is best done in the time dependent form (with $t$ implicit in functions).  Writing out the right hand side one has  
\begin{equation}\label {eig2}
\left[{\begin{array}{cc}
\partial \psi_{\alpha}(xH)/\partial t \\
\partial \psi_{a}(xH)/\partial t
\end{array}}\right]=
\left[{\begin{array}{cc}
{\mathfrak{h}_{\alpha}}\psi_{\alpha}(xH) -\mathfrak{p}\psi_{a}(xH) \\
{\mathfrak{p}^*}\psi_{\alpha}(xH)+ \mathfrak{h}_{a}\psi_{a}(xH)
\end{array}}\right].
\end{equation}
The right hand side is just eq. (\ref{transitions}) again.  At an instant of observation, as registered by the change in the wave function $\partial \psi_{a}(xH)/\partial t$ of the \emph{surroundings}, the system $\psi_{\alpha}(xH)$ reports its \emph {present} state via the operator $\mathfrak{p}^\ast$.  However, since the system is in contact with its surroundings \emph {via} the $\mathfrak{p}\psi_{a}(xH)$ term, its state will evolve and the next observation will find the system in a different state.   Whether the change of state is large or small depends on the strength of the coupling, and that requires formulation and solution of a specific problem.  In any event, the physical content of eq. (\ref {eig2}) is that the properties of matter are observed through an interaction with experimental apparatus (or by personal observation). 

The $\mathfrak {h}_\alpha$ operator in eq. (\ref{eig2}), restricted to a single particle, is a spin operator that is a function of the coordinates in the (Grassmannian) cross section as is seen in eq. (\ref{h_op}). This operator is involved because the Lie derivative transports the system in a tangent plane, and this projects onto a motion along the $Sp\left(1\right)$ fiber because the cross section is not flat.  The block diagonal $\mathfrak {h}$ angular momentum operators have a clear interpretation as internal degrees of freedom, but their external influence is conveyed in interactions through $\mathfrak p$, as is clear in eqs. (\ref{p_op2}-\ref{p_op1}). 
 
The $(\delta_{\alpha\beta}+\zeta_{\alpha b}\bar \zeta_{\beta b})$ and $(\delta_{ab}+\bar \zeta_{\mu a}\zeta_{\mu b})$ terms in eqs. (\ref{p_op2}-\ref{p_op1}) reduce to the identity for small $|q|$ and otherwise convey the non-euclidean character of the space.  The $q_{\alpha b}h_{ab}$ and  $q_{\mu a}h_{\alpha \mu}$ terms are analogous to distance$\times$angular momenta; the implications of the non-linear parts of $\mathfrak p$ will be discussed elsewhere.

\section*{The Electromagnetic Field from the Lie Algebra}

Use eq. (\ref{eqmot}) in eq. (\ref {eig2}) and write the latter as 
\begin{equation}\label {op1}
\left[{\begin{array}{cc}
\partial \sigma_{1}\psi_{1}(x)/\partial t \\
\partial \sigma_{2}\psi_{2}(x)/\partial t
\end{array}}\right]=
\left[{\begin{array}{cc}
{\mathfrak{h}_{1}}\sigma_{1}\psi_{1}(x) -\mathfrak{p}\sigma_{2}\psi_{2}(x) \\
{\mathfrak{p}^*}\sigma_{1}\psi_{1}(x)+ \mathfrak{h}_{2}\sigma_{2}\psi_{2}(x)
\end{array}}\right].
\end{equation}
The indices have been changed to make the equations easier to read for the case of just two particles.  Here $\sigma_{k}=\sigma(H^{-1}_{k}), k={{1,2}}$.  The unitary representations, $\sigma_{k}$, of the isotropy subgroups commute with the operators, as they are not functions of $x$, the Grassmannian coordinates.  However, they might be functions of the time $t$.  Collect terms to yield 
\begin{equation}\label {op1}
\left[{\begin{array}{cc}
\sigma^{-1}_{2}[\partial \sigma_{1}\psi_{1}(x)/\partial t- \sigma_{1}\mathfrak{h}_{1}\psi_{1}(x)]\\
\sigma^{-1}_{1}[\partial \sigma_{2}\psi_{2}(x)/\partial t-\sigma_{2}\mathfrak{h}_{2}\psi_{2}(x)]
\end{array}}\right]=
\left[{\begin{array}{cc}
\mathfrak{o}_{1} \psi_{1}(x) \\
\mathfrak{o}_{2}\psi_{2}(x)
\end{array}}\right]=\left[{\begin{array}{cc}
 -\mathfrak{p}\psi_{2}(x) \\
{\mathfrak{p}^*}\psi_{1}(x)
\end{array}}\right],
\end{equation}
where the $\mathfrak {o}_k$ are the operators defined by the left-most members of the equation.  The off-diagonal components of the infinitesimal generators, eq. (\ref{Hop}), of the Lie algebra are $p_{\alpha a}=\bar\partial_{\alpha a}+ \zeta_{\alpha b}\zeta_{\mu a}\partial_{\alpha b}$ and their conjugates.  Near the origin, \emph{i.e.}, all $\zeta_{\alpha b}$ small, the generator is just $p_{\alpha a}=\bar\partial_{\alpha a}$.  For this section, the non-linear terms in the generators will be suppressed.   Since we will be dealing with only one derivative, it is convenient to revert to the quaternion form of the differential operator, and to define the $\psi_{k}(x)\in \mathbb{H}$.  It is also convenient to make use of scalar-vector notation, so that $p=\partial_{0}{\bf e}-\partial_{1}{\bf i}-\partial_{2}{\bf j}-\partial_{3}{\bf k}=\partial_{0}-\nabla$, where $\partial_k=\partial/\partial x_k$ and the basis element ${\bf e}$ is implicit in the scalar term. 

The product of two quaternions, $v = v_0+\bf v$ and $w=w_0+\bf w$, written in the scalar-vector notation is $vw = (v_{0}w_{0}-{\bf v}\cdot{\bf w})+(v_{0}{\bf w}+w_{0}{\bf v}+{\bf v}\times{\bf w})$. The identity term, with its Lorentz signature, makes another contact with Special Relativity.  Define $\psi_1 = A_{0}{\bf e}+A_{1}{\bf i}+A_{2}{\bf j}+A_{3}{\bf k}=A_0 + {\bf A}$ in suggestive notation.  The action of $\mathfrak {p}^\ast$ on $\psi_1$ is 
\begin{align*}
p^{\ast}\psi_1={}& (\partial_{0}+\nabla)(A_0 + {\bf A})=(A_{0,0}-\nabla\cdot{\bf A})+({\bf A}_{,0}+\nabla A_0) + \nabla\times{\bf A}\\
={}& (A_{0,0}-\nabla\cdot{\bf A})-{\bf E}+{\bf B},
\end{align*}
where $f_{,0}=\partial_{0}f$.  The source term, $\psi_1$, acts as the electromagnetic vector potential in the linear regime, as $\bf E=-{\bf A}_{,0}-\nabla A_0$ and $\bf B=\nabla\times{\bf A}$ are clearly defined as the electric and magnetic fields, respectively.  Note that the time derivative is with respect to the cyclic variable.  The scalar term, $A_{0,0}-\nabla\cdot{\bf A}$, has no analogue in Maxwell's equations.    It is conjectured that the non-linear part of $\mathfrak p$ represents the action of heavy bosons.  

An alternative to this presentation is to perform an `anti-Wick' rotation on the identity components of $A$ and $\mathfrak{p}$, sending them to $\tilde{A}$ and $\mathfrak{\tilde p}$, and calculate as above.  (This is equivalent to giving the quaternions a Lorentz signature.)  This also gives Maxwell's equations with electric and magnetic fields being real and imaginary components.  Application of $\mathfrak{\tilde{p}}^\ast$ to $\mathfrak{\tilde p}A$ gives Maxwell's equations as real and imaginary components of the action of $\mathfrak{\tilde p}^\ast$, and the identity component of $\mathfrak{\tilde p}\tilde A$ acts as the source.  The introduction of $\sqrt{-1}$  is not advocated in general, as this introduces bi-quaternions into the algebra; at present there is no obvious need for this extension.  Adler\cite{Adler} gives a more substantial reason for avoiding bi-quaternions.

It will be interesting to see if the three types of terms in the quaternion product -- scalar, scalar-vector, and vector -- can be mapped into electromagnetic, weak, and strong interactions, respectively.  The Lorentz signature of the scalar term has a clear connection with the Special Theory and electromagnetism, so that is a start.  The additional clue to this connection is provided by the four bosons -- $\gamma, W^{\pm}, \textrm{and } Z$ -- which might map to the quaternion basis.  The structure of the theory is fundamentally about multiplication of  particle states by matrix elements, and these represent contact terms between bosons and fermions.
 
\section*{The Alternative Representation of $Sp\left(n\right)$}

To this point the theory has been developed in the representation $U\left(n,\mathbb{H}\right)$.  Given the fact that quaternions are not commutative, it may be more convenient for some calculations to work in the $Sp\left(2n,\mathbb{C}\right)$ version.  The conjugation operation $q^*=j^{-1}q^{\prime}j$ provides just what is needed to map between the two representations.\cite{Varad}  For $g\in U\left(n,\mathbb{H}\right)$ we have $g^*g = J_n^{-1}g^{\prime}J_n g = 1$, so that $g^{\prime}J_n g = J_n$, where $J_n:=\mathbf{1}_n\bigotimes j$ as before.  Now, there exists a permutation $\mathcal{P}$ such that $\mathcal{P}: \left(\mathbf{1}_n\bigotimes j\right) \to j\bigotimes \mathbf{1}_n$, and acting on $g$ gives a permuted form $\mathcal{P}: g \to \mathcal{G}$.  For $\mathcal{J}=j\bigotimes \mathbf{1}_n$, this gives $\mathcal{G^{\prime}JG}=\mathcal{J}$, which is the standard definition of the symplectic group over the complex numbers [$\mathcal{G} \in Sp\left(2n,\mathbb{C}\right)$]; the group preserves a complex skew-symmetric bilinear form.  But the group is also unitary, \emph{i.e.}, $Sp\left(n\right)\sim Sp\left(2n,\mathbb{C}\right)\cap U\left(2n,\mathbb{C}\right)$, as noted above, so that we also have $\mathcal{G^*G}=1$.  These two properties yield the Lie algebra in this representation in the form
\[
\mathfrak{g}\sim \left[{\begin{array}{cc}
\mathfrak{a} & \mathfrak{b} \\
{-\mathfrak{b}^*} & {-\mathfrak{a}^{\prime}}
\end{array}}\right]; \quad \mathfrak{a}^* = -\mathfrak{a}; \quad \mathfrak{b}^{\prime} = \mathfrak{b}
\]
where $\mathfrak{a}$ and $\mathfrak{b}$ are complex matrices.  Cosets in this representation are messier to work with than those in the quaternion basis, so this representation is not pursued further here.  For example, the components of the $\mathfrak{sp}(1)$ fibres are split between the diagonal elements of $\mathfrak a$ and $\mathfrak b$.  In any event, one should note that the $\mathfrak{a}$-sector of this representation, taken alone, is isomorphic to $\mathfrak{u}(n)$.  

\section* {$S^4$ and the Conformal Group} 

The isomorphism $Sp\left(2\right)/Sp\left(1\right)\times Sp\left(1\right) \sim SO\left(5\right)/SO\left(4\right) \sim S^4$ enables one to use real coordinates for calculations.  In the $\mathbb R$ ring the manifold $S^4$ is defined by
\[
\sum\limits_{i=0}^4{x_i^2}=1.
\]
$SO\left(5\right)$ acts by linear fractional 
transformations on the inhomogeneous coordinates $y_k=x_k/x_0, 1\le k \le 4$, with $1+yy^{\prime}=1/x_0^2 \ge 1$.  Evaluation of the invariant metric is a standard calculation, yielding the line element in the $y$-coordinates as the Fubini-Study metric\cite{K&N}
\begin{equation}\label{s4}
ds^2 = \left(1+yy^{\prime}\right)^{-1}dy\left(1+y^{\prime}y\right)^{-1}dy^{\prime}.
\end{equation}
The relation between the metrics on $Sp\left(2\right)/Sp\left(1\right)\times Sp\left(1\right)$ and $SO\left(5\right)/SO\left(4\right)$ is most clearly seen on pulling both metrics back to the sphere.  The self-dual and anti-self-dual connections, Yang-Mills action, and instantons are more accessible from the quaternion version,\cite{At,Lawson} but the real version exposes the relation with the de Sitter space. 

The 4-sphere is the surface of the 5-ball, $B^5$, and the exterior of the sphere is obtained by inverting the ball through the $S^4$ surface.  The extended Lorentz group $SO\left(1,5\right)$ acts on the ball (an anti-de Sitter, AdS, space), while the de Sitter  space ($vv^{\prime}-v_0^2 = 1 \Rightarrow xx^{\prime}-1>0;x=v/v_0$) is the inverted ball (here $v$ is a 5-dimensional vector).   The group $SO\left(1,5\right)$ is the conformal group of $S^4$\cite {Lawson,Mal,Chu}.  The boundary, $S^4$, of either $B^5$ or its inverse is approached in the limit $v_0 \to \infty$. 

\section* {Charge and Characteristic Classes}
Gauss's theorem in Newtonian space defines charge as an integral of a vector field (the electric field) over a closed surface of $\mathbb{R}^3$.  This is a topological definition of charge.  Since the boundary of $S^4$ is empty, we have to resort to a different topological invariant, and the Euler characteristic, $\chi(M)$ for a manifold $M$, seems appropriate.\cite{Milnor, Milnor2}  For the 4-sphere, $\chi(S^4)=2$.   Since $S^4$ has a smooth vector field that vanishes at the poles, the poles are singularities.  The poles have all the hallmarks of positive and negative charges, as suggested by Fig. (\ref{fig:S4}). 
\begin{figure}[htbp]  \begin{center}
\includegraphics[width=8cm]{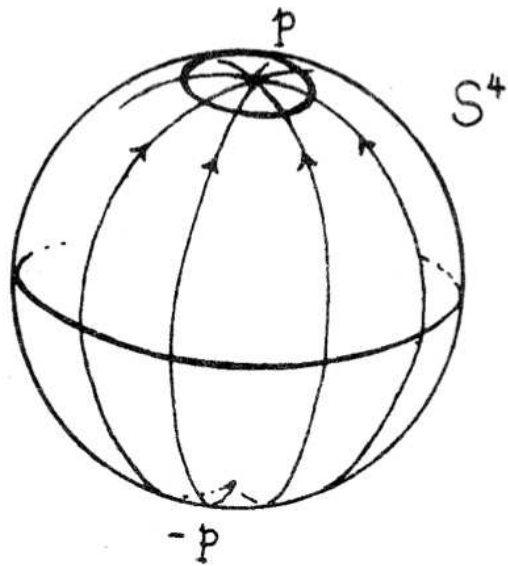}
\end{center}
\caption{The space $S^4$ of two particles and accompanying instantons. Reprinted from Figure 10 (p. 14) in H. Blaine Lawson, Jr. \emph{The Theory of Gauge Fields in Four Dimensions}, CBMS Regional Conference Series in Mathematics Volume 58 (Providence: American Mathematical Society, 1985). \copyright 1985 by the American Mathematical Society.}\label{fig:S4}
\end{figure}
For a single particle among many, \emph{i.e.}, $Sp(n+1)/Sp(1)\times Sp(n)=S^{4n}$, the Euler characteristic is the same: $\chi(S^{4n})=2$.  For large $n$ this is equivalent to a charge and its image charge in classical statics.  Neutrinos do not fit into this picture because they do not appear as stationary states; they are associated with energy exchange.  

All Grassmannians, $Sp(n+k)/Sp(k)\times Sp(n)$, of interest here have real dimension $4kn$ and so have non-trivial Pontrjagin numbers.  In particular, see Theorem 16.8 of Milnor and Stasheff\cite{Milnor2} in relation to eq. (\ref{product}) above.  The homotopy groups of $Sp(n)$ are well known.\cite{Bott} The study of characteristic classes for the geometric structures that are being discussed likely will be important in several ways.  Besides charge, the use of characteristic classes to study the intersection in the bundle spaces should be fruitful, as intersections have all the hallmarks of level crossings in quantum theory, which can be associated to energy transfer.  Intersections also appear to be related to  ``bubbling-off"\cite{Lawson,Mal2}, as mentioned above.  Bubbling-off would appear to be particle creation; the converse is particle annihilation.  Both phenomena might be manifestations of the dynamics of high-dimensional representations.  All of these topics will require significant development.

Reflecting on the action of $\mathfrak{p}$ and $\mathfrak{p}^\ast$ on $\psi_1$ and $\psi_2$, and the fact that $\mathfrak{p}$ and $-\mathfrak{p}^\ast$ differ in the sign of the identity (commuting) component, sheds further light on the identity component as a cyclic time coordinate.  A clock that is running in the clock-wise (longitudal) direction when seen from the south pole of $S^4$ will be running in the counter-clock-wise direction when viewed from the north pole.    This cyclic-time reversal symmetry permeates everything in the symplectic group.  As discussed above, the time order of events is determined by Galilean (winding-number) time, not the cyclic variables.

\section* {Mass and Curvature}

Having shown that spin is defined by the group algebra and charge by topology, it remains to discuss mass. The Newtonian concept of inertial mass arises in the context of motion relative to an inertial frame.  In BL relativity, the motion of an object is always relative to other objects, so the concept of inertial mass does not arise.  An object always has internal motions, but it moves relative to other objects only by virtue of interactions with other objects, and these  interactions are conveyed by the curvature tensor.  The classical definition of mass is the resistance to motion on application of a force, and since force has been identified with curvature, curvature has to provide a measure of mass and gravity.  Since all motions in compact spaces are intrinsically cyclic, gravity can be interpreted as the fictitious macroscopic force that is imposed on a Euclidean space to convert linear motion into cyclic motion.

In the general $Sp(n)/Sp(k)\times Sp(n-k)$ setting, the trace and determinant of the curvature tensor qualify as appropriate invariants that will provide the simplest scalar functions via a mapping: $\Omega \to \mathbb{R}$, uniformly for all $k$ and very large $n$.  

To evaluate trace and determinant of the curvature = metric tensor requires some elementary calculations.  Let $dX$ be a $k\times n$ matrix with conjugate transpose $dX^\ast$.  The row form of $dX$ will be denoted by $dx$, where $dx=(dX_{11}, dX_{12},\cdots,dX_{1n}, dX_{21},\cdots,dX_{kn})$.  It is easy to show that $\textrm{tr}(AdXBdX^\ast)=dx(A'\otimes B)dx^\ast$.  Here $A$ and $B$ are conformable with $X$, $A'$ is the transpose of $A$, and $A'\otimes B$ is the direct product of matrices $A'$ and $B$.  The trace and determinant are easily shown to be $\textrm{tr}(A'\otimes B)=\textrm{tr}(A)\textrm{tr}(B)$ and $\textrm{det}(A'\otimes B)=[\textrm{det}(A)]^{\textrm{dim}(B)}[\textrm{det}(B)]^{\textrm{dim}(A)}$, since $\textrm{tr}(A')=\textrm{tr}(A)$ and $\textrm{det}(A')=\textrm{det}(A)$.

The trace of the metric tensor is simply obtained from $(1+Q^\ast Q)^{-1}=1-Q^\ast(1+QQ^\ast)^{-1}Q$.  It follows that $\textrm{tr}[(1+Q^\ast Q)^{-1}]=2(n-k)+\textrm{tr}[(1-(1+QQ^\ast)^{-1}QQ^\ast]=2(n-k)+\textrm{tr}[(1+QQ^\ast)^{-1}]$.  The presence of $n-k$ in the trace measure of curvature is unattractive in the calculation of an integral invariant, suggesting that the determinant of the curvature tensor is a more pleasing quantity.  The determinant of the metric tensor is formally
\[
\textrm{det}(\Omega)=g=\textrm{det}[(1+QQ^\ast)^{-1}\otimes (1+Q^\ast Q)^{-1}]=[\textrm{det}(1+QQ^\ast)]^{-(k+n)}=|1+QQ^\ast|^{-(k+n)},
\]
but which requires care in converting to real variables.  The integral of the Gaussian curvature over the normalized volume,  
\[
\int\textrm{det}(\Omega)\sqrt{g}dQ/\int\sqrt{g}dQ,
\]
is not a trivial calculation for $k>1$.  The determinants have to be expressed in terms of the eigenvalues of $1+QQ^\ast$, as there is no satisfactory expression for the determinant of a general quaternionic matrix.  Fortunately, $QQ^\ast$ is `hyper-Hermitean' and has real eigenvalues.  In converting the volume element to polar coordinates, the matrix of eigenvalues has a normalizer with $3k$ real dimensions.  Experience with this class of matrices\cite{Mehta} shows that their polar volume elements contain terms of the type $\Pi_{i<j}|\lambda_{i} - \lambda_{j}|^{\beta}$ where the $\lambda_i$ are eigenvalues of $QQ^\ast$.  The integrals are related to Selberg's integral,\cite{Mehta,Sel} and are left to another time.

\section* {Representations of $Sp(k)$ for Small $k$}

The primary reason for being interested in $Sp(n)/Sp(k)\times Sp(n-k)$ for small $k$ and large $n$ is that these spaces presumably provide insight into elementary particle structure.  The connection is stated as a conjecture: 

\begin{itemize}
\item Leptons are represented by $Sp(1)\times \psi_1[Sp(n)/Sp(1)\times Sp(n-1)]$
\item Mesons by $Sp(2)\times \psi_2[Sp(n)/Sp(2)\times Sp(n-2)]$
\item Baryons by $Sp(3)\times \psi_3[Sp(n)/Sp(3)\times Sp(n-3)]$
\end{itemize}
The conjecture is stated so as to make the internal symmetries clearly identifiable.  The relation between the rank of these spaces and topology is palpable.  If this conjecture is correct, it is also clear why individual quarks are not isolable:   A higher dimensional representation of, say, $Sp(3)$ might ``come apart" (through its interactions with the surroundings) into pieces that are classified in $Sp(1), Sp(2)$, or $Sp(3)$; \emph{i.e.}, the decay products are either leptons, mesons or baryons.  The \emph{constituents} of a representation of $Sp(3)$ do not have an independent existence.  

As has been emphasized repeatedly, an isolated system is described by an irreducible representation of $Sp(n)$ of the appropriate dimension $n$.  There are various ways that representations might be constructed -- each route will entail extensive calculations.  Analytical representations can be obtained by solution of the first order differential equations provided by the generators of the Lie algebra that are presented above.  The solution of the Laplace-Beltrami operator on $S^4$ will provide one aspect of these solutions, as will be seen later.  The second approach to representations is direct construction from the matrix representations, $g=xH$.  Direct products $g\otimes g=(x\otimes x)(H\otimes H)$ will give, in general, reducible representations which might then be rendered into irreducible components.  The method of induced representations\cite{Folland} provides yet another route.

For present purposes, the geometry of the root space of $\mathfrak{sp}(n)$ may classify the elementary particles.  Fulton and Harris\cite{Fulton} develop the representations for $\mathfrak{sp}_{n}\mathbb H$ in the $\mathfrak{sp}_{2n}\mathbb C$ setting, where they show that the roots of the Lie algebra $\mathfrak{sp}(n)$ are vectors $\pm L_i \pm L_j, 1\le i\le j\le n$.  The root space for $\mathfrak{sp}(3)$ is shown in Fig. (\ref{fig:sp3}).  

\begin{figure}[htbp]
\begin{center}
\includegraphics[width=8cm]{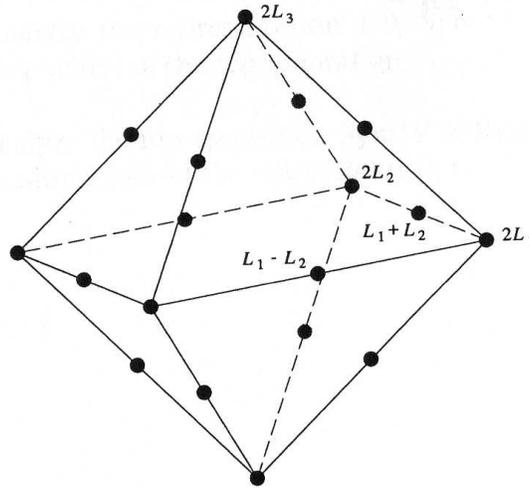}
\end{center}
\caption{The root space for $\mathfrak{sp}(3)$.  Reprinted from a figure on p. 254 in William Fulton and Joe Harris \emph{Representation Theory -- A First Course} Graduate Texts in Mathematics Vol. 129 (New York:  Springer-Verlag, 1991).  Used by permission.  Graphic created by Chandler Fulton.}
\label{fig:sp3}
\end{figure}

Particle assignments conforming to the conjectures presented here can be quickly seen to be similar to the standard $\mathfrak{su}(n)$ assignments.  Since the roots of $\mathfrak{sp}(k-1)$ span a subspace of  $\mathfrak{sp}(k)$, quark assignments from the $\mathfrak{sp}(3)$ and $\mathfrak{sp}(4)$ root spaces will apply to mesons in $\mathfrak{sp}(2)$ in just the same way as is done for the assignments with $\mathfrak{su}_3$ and $\mathfrak{su}_4$.  

Leptons correspond to $\psi_1\sim \pm 2L_i$, \emph{i.e.}, one-dimensional subspaces.  Meson states will be represented as linear combinations of two weight vectors $L_i$ with integer coefficients.  Such a state might be represented by 

\[
\psi_2(x)=\left[\begin{array}{ccc}
  \psi_{a}(x)\\
  \psi_{b}(x) \\
\end{array}
\right]\sim\left[\begin{array}{ccc}
  L_1+L_2\\
  L_1-L_2 \\
\end{array}
\right]
\]
for example.  To conform to current particle assignments, let $L_1=$u, $L_2=$d, $L_3$=s, \emph{etc.}, where u, d, s, c are the familiar quark assignments.  The octahedral symmetry of the root space allows families to be grouped in fours, $\pm L_i \pm L_j$, \emph{e.g.}, $L_1+L_2=$ud, $-L_1+L_2=\bar{\textrm{u}}$d, \emph{etc.}, and as well as threes and sixes.  Continuing in this vein, baryons will be linear combinations of three weight vectors.  It should also be noted that the root space is only a part of the assignment.  The quaternionic variables provide three additional degrees of freedom, the basis vectors $\bf{i,j,k}$, so that a state such as $L_1{\bf i}+ L_1{\bf j}+ L_2{\bf k}$, similar to the state assignment (uud) for the proton, can be made without violating the exclusion principle.  The quaternion basis is equivalent to the color degree of freedom in quantum chromodynamics. 

It remains to be seen if all meson and baryon assignments can be made consistent in the $\mathfrak{sp}(n)$ root space, but since $\mathfrak{su}(n)$ is a subspace of $\mathfrak{sp}(n)$, the larger algebra is at least compatible with the currently accepted assignments.  Furthermore, $\mathfrak{su}(n)$ may provide a route to understanding groupings into families of four that have not otherwise been explained.  Having pointed out the salient relations between $\mathfrak{su}(n)$ and $\mathfrak{sp}(n)$, the correspondence between the assignments in the two algebras remains to be explored.

\section*{$Sp_{2}/Sp_1^2$ Once Again}
The purpose of this section is to discuss a few additional aspects of the theory in the context of an illustrative calculation.  The simplest equation is provided by the action of the Laplace-Beltrami operator on scalar functions on $S^4\sim Sp_{2}/Sp_1^2$.  The metric is $ds^2=(1+qq^*)^{-1}dq(1+q^*q)^{-1}dq^*=(1+|q|^2)^{-2}dqdq^*$.  The substitution $q=\cot(\omega/2)u$, with $uu^*=1$ gives $ds^2=d\omega^2\mathbf{1}+(\sin\omega)^2\delta u\delta u^*$, where $\delta u=duu^*=-udu^*=-\delta u^*$.  (An uninteresting numerical factor was dropped.) The reason for choosing $\cot(\omega/2)$ rather than $\tan(\omega/2)$ is that we want a singularity at the origin, as will be explained shortly.  A convenient parameterization of $Sp_1$ is 
\begin{equation*}
u=\exp (\mathbf{k}\beta/2)\exp (\mathbf{i}\alpha/2)\exp (\mathbf{k}\gamma/2)=bac
\end{equation*}
such that 
\begin{equation*}
\delta u = (1/2)b[d\alpha\mathbf{i}+d\beta\mathbf{k}+d\gamma\exp(\alpha\mathbf{i})\mathbf{k}]b^*
\end{equation*}
from which it follows that
\begin{equation*}
ds^2=4d\omega^2+\sin^{2}\omega[d\alpha^2+d\beta^2+d\gamma^2+2(\cos\alpha) d\beta d\gamma].
\end{equation*}
The factor of four in this equation plays an interesting role, as will now be seen.  The LB operator, $\Delta$, on $S^4$ with this parameterization is 
\begin{equation*}
\Delta=\partial^2_\omega+3\cot\omega\partial_\omega+(4/\sin^{2}\omega)[\partial^{2}_\alpha+\cot\alpha\partial_\alpha+(1/\sin^{2}\alpha)(\partial^{2}_\beta+\partial^{2}_\gamma-2\cos\alpha\partial_\beta\partial_\gamma)].
\end{equation*}
The operator in brackets is the total angular momentum operator having the usual solutions, leaving 
\begin{equation}\label{LBonS4}
\Delta f(\omega)=f^{\prime\prime}+3\cot\omega f^{\prime}-[2\ell(2\ell+2)/ \sin^{2}\omega]f=\delta(\omega)
\end{equation}
where $f^{\prime}=\partial f/\partial \omega$.
The factor of four in the $S^3$ angular momentum operator is important in two ways: (\emph{i}) by allowing half-integer spin while permitting a polynomial solution to the $\omega$ equation (as will be seen), and (\emph{ii}) by embedding the $S^3$ solutions into $S^4$, which is the role of the $+2$ in $2\ell(2\ell+2)$.  

The delta function on the right in eq. (\ref{LBonS4}) represents a source at the pole, and this requires comment.  The metric in eq. (\ref{s4}) is for the antipodal projection of $S^4$ onto $\mathbb{R}^4$.  In the $Sp_{2}/Sp^2_1$ picture, symmetry recommends that the two particles be placed at antipodal points of $S^4$; these two points map to $q=0$ in the projection [see Fig. (\ref{fig:S4}].  Particles are singularities in otherwise smooth manifolds; the space-time `between' particles is smooth. If one only looks for smooth solutions of  eq. (\ref{LBonS4}) there will be no distinguished points.  The singularity at $\omega=0$ is a statement of these general principles. In topological language, physical singularities are `holes'.   

The solution of eq. (\ref{LBonS4} for $\ell=0$ is $f_0=-\cot\omega/\sin\omega+\ln[\tan(\omega/2)]$.  This static and integrable potential is continuous at $\omega=\pi/2$; continuity at the equator is essential to ensure continuity of the field on $S^4$ except at the poles.  On attempting to solve the equation for $\ell \neq 0$ a singularity as $q\to\infty$ can be avoided by adding $\theta^{2}f_\ell$ to the left in eq. (\ref{LBonS4}), which is equivalent to converting the operator to $(\Delta -\partial^{2}_{t})f_\ell$, with $f_\ell=g_{\ell}\exp{(i\theta t)}$.  One of the general solutions is
\[
g_\ell=(\sin\omega)^{-2(\ell+1)}\sum\limits_{n=0}^N { a_{n}(\sin\omega)^{2n}}
\]
 with 
\[
a_{n} = \frac{(2\ell -n)!(N-2\ell-3/2+n)!}
{n!(N-n)!}
\]
and $\theta=\sqrt{(\ell+1-N)(\ell-1/2-N)}$.  The polynomial terminates at $n=N$, where $N<\ell+1$ if $\ell$ is an integer and $N<\ell-1/2$ if $\ell$ is a half-integer. The solution is again singular at the origin, but the time dependence insures that it is continuous at $\omega=\pi/2$.  The integer solutions should represent the potentials for bosons that mediate the interaction between the two fermions in a meson.  Half-integer solutions might be related to neutrinos.  

The ground state and excited states of a bare meson are encompassed in this solution.  The meson is bare because interactions with the surroundings are encompassed by $Sp(n)/Sp(2)\times Sp(n-2)$; the solutions to eq. (\ref{LBonS4}) presumably give a part of the energy, but interactions with the surroundings are required to get the total energy.  An interesting aspect of $g_\ell$ is that it is not integrable for $\ell>1/2$, which may have some bearing on questions of stability.  
   
 \section* {Discussion and Further Interpretation}
The space between two particles, taken in isolation, is the rank-one sphere, $S^4$, which is a smooth manifold that carries a representation of the interaction between particles.  This development of Berkeley's and Leibniz's notions of the relativity of location and motion leads to the remarkable conclusion that interactions create spacetime -- spacetime has no objective definition apart from its identification as the manifold of interactions between physical objects.  Those of us trained to think of physical phenomena as occurring in an infinite flat space may find this concept difficult to accept.   However, some thought will convince one that this empirical definition of spacetime is fully compatible with reality.  An empty, flat Euclidean space is an ideal that cannot be observed.  One cannot determine the character of physical space except by observation, but to observe anything requires the presence of matter, and more particularly of interactions between two or more material objects (observers included).  There is no operational definition than can be given to space except through such an interaction.\footnote{In a clear case of anticipatory plagiarism, Kant observed that ``It is easily proved that there would be no space and no extension if substances had no force to act outside themselves.  For without a force of this kind there is no connexion, without this connexion no order, and without order no space."\cite{Alexander}} Boundless Euclidean spaces can be imagined, as can more bizarre mathematical spaces of any dimension, but if they cannot be observed they have no physical reality.  Furthermore, it is impossible to observe a non-compact space.  Given the finite velocity of signals, all interactions between material objects are necessarily restricted to compact domains.  

Geodesics on the sphere $S^4$ are all paths from pole to pole.  The coset space is parameterized by 
\[
\exp\left[
\begin{array}{cc}
 0   &q   \\
 -q^\ast  &0   
\end{array}
\right]
\]
where $q$ is an unrestricted quaternion.  On extracting a modulus $\omega t$, where $t$ is Galilean time and $\omega$ is a frequency so as to parameterize the geodesics, this becomes 
\[
\exp\left[\omega t\left(
\begin{array}{cc}
 0   &u   \\
 -u^\ast  &0   
\end{array}
\right)\right]=\left[\begin{array}{cc}
\cos(\omega t)\bf 1 & \sin(\omega t)u\\
-\sin(\omega t)u^\ast & \cos(\omega t)\bf 1\end{array}\right]
\]
where $u$ is a unit quaternion.  The frequency in this equation is a candidate for the energy of the system, and its magnitude is to be found in the solutions sketched in the previous section.  The `bare' system energies computed by this means are probably not as interesting as the quantities that might be calculated from consideration of the coupling of systems to surroundings.  

A matrix in $Sp(n)$ that represents interactions is constructible from eigenvectors and eigenvalues, which clearly provides the relation between bosonic interactions and fermionic eigenvectors.  The gauge (structure) group, $Sp^{n}_1$, acts on the vector space of fermion states by $Sp^{n}_1: \Psi\to \sigma_{k}(Sp_{1})\psi_{k}, 1\le k\le n$ in accordance with eq. (\ref{induce}).  The distinction between a representation of $Sp_1$ and the fermion on which it acts is in the normalization (\emph{weight}) of the fermion, which is determined by its relation to other particles.  

\bigskip
\textsc{Acknowledgment}

The author is grateful for the hospitality of Profs. Ulrich Suter and Peter G\"{u}nter at the ETH during the Spring of 2008.  A portion of this work was accomplished at that time.  Profs. John Sullivan, Gerald Folland, and Sara Billey at the University of Washington provided answers to many mathematical questions.

\bibliographystyle{aip}

\end{document}